\documentclass[graybox]{svmult}

\usepackage{type1cm}        % activate if the above 3 fonts are
\usepackage{makeidx}         % allows index generation
\usepackage{graphicx}        % standard LaTeX graphics tool
\usepackage{multicol}        % used for the two-column index
\usepackage[bottom]{footmisc}% places footnotes at page bottom

\usepackage{newtxtext}       % 
\usepackage{newtxmath}       % selects Times Roman as basic font
\usepackage{aas_macros} %needed to define journal macros generated by 

\makeindex             % used for the subject index
\begin{document}

\title*{Diagnostics from solar and stellar glitches}
% Use \titlerunning{Short Title} for an abbreviated version of
% your contribution title if the original one is too long
\author{Margarida S. Cunha}
% Use \authorrunning{Short Title} for an abbreviated version of
% your contribution title if the original one is too long
\institute{Margarida S. Cunha \at Instituto de Astrof\'isica e Ci\^encias do Espa\c{c}o, Universidade do Porto, CAUP, Rua das Estrelas, 4150-762 Porto, Portugal and School of Physics and Astronomy, University of Birmingham,Birmingham, B15 2TT, United Kingdom \email{mcunha@astro.up.pt}}
%
% Use the package "url.sty" to avoid
% problems with special characters
% used in your e-mail or web address
%
\maketitle

\abstract{Sudden changes in the internal structure of stars, placed at the interface between convective and radiative regions, regions of partial ionisation, or between layers that have acquired different chemical composition as a result of nuclear burning, often produce specific signatures in the stars’ oscillation spectra. Through the study of these signatures one may gain information on the physical processes that  shape the regions that produce them, including diffusion and chemical mixing beyond the convectively unstable regions, as well as information about the helium content of stars. In this talk, I will review important theoretical and observational efforts conducted over the years towards this goal. I will emphasise the potential offered by the study of acoustic, gravity, and mixed modes observed in stars of different mass and  evolutionary stages, at a time when space-based data is allowing us to build on the knowledge gained from the study of the sun and white dwarfs, where these efforts have long been undertaken, extending the methods developed to stars across the HR diagramme.}

\section{Glitch, glitch signature, and mode trapping}
\begin{figure}[t]
\includegraphics[scale=.35]{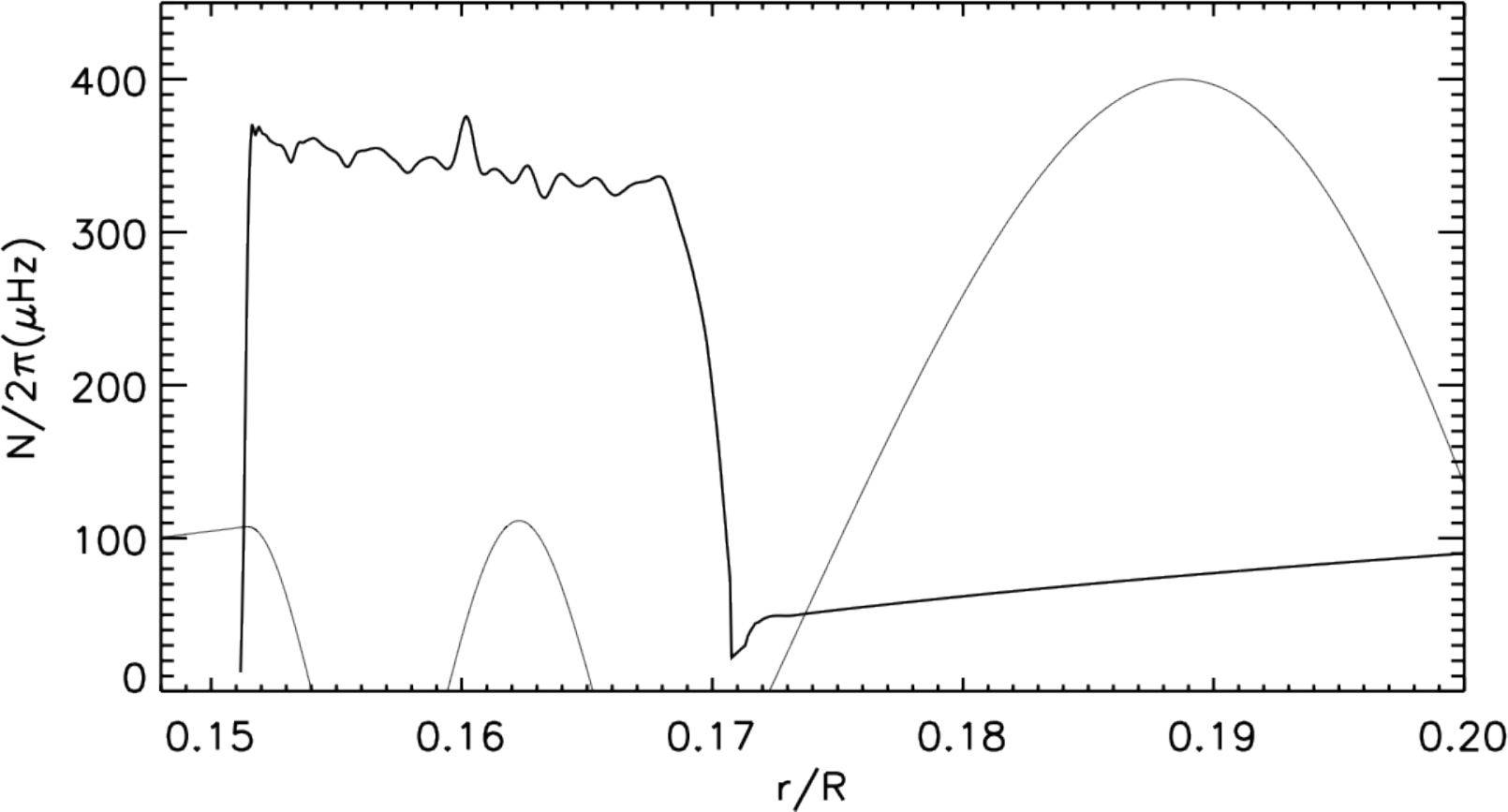}
\includegraphics[scale=.35]{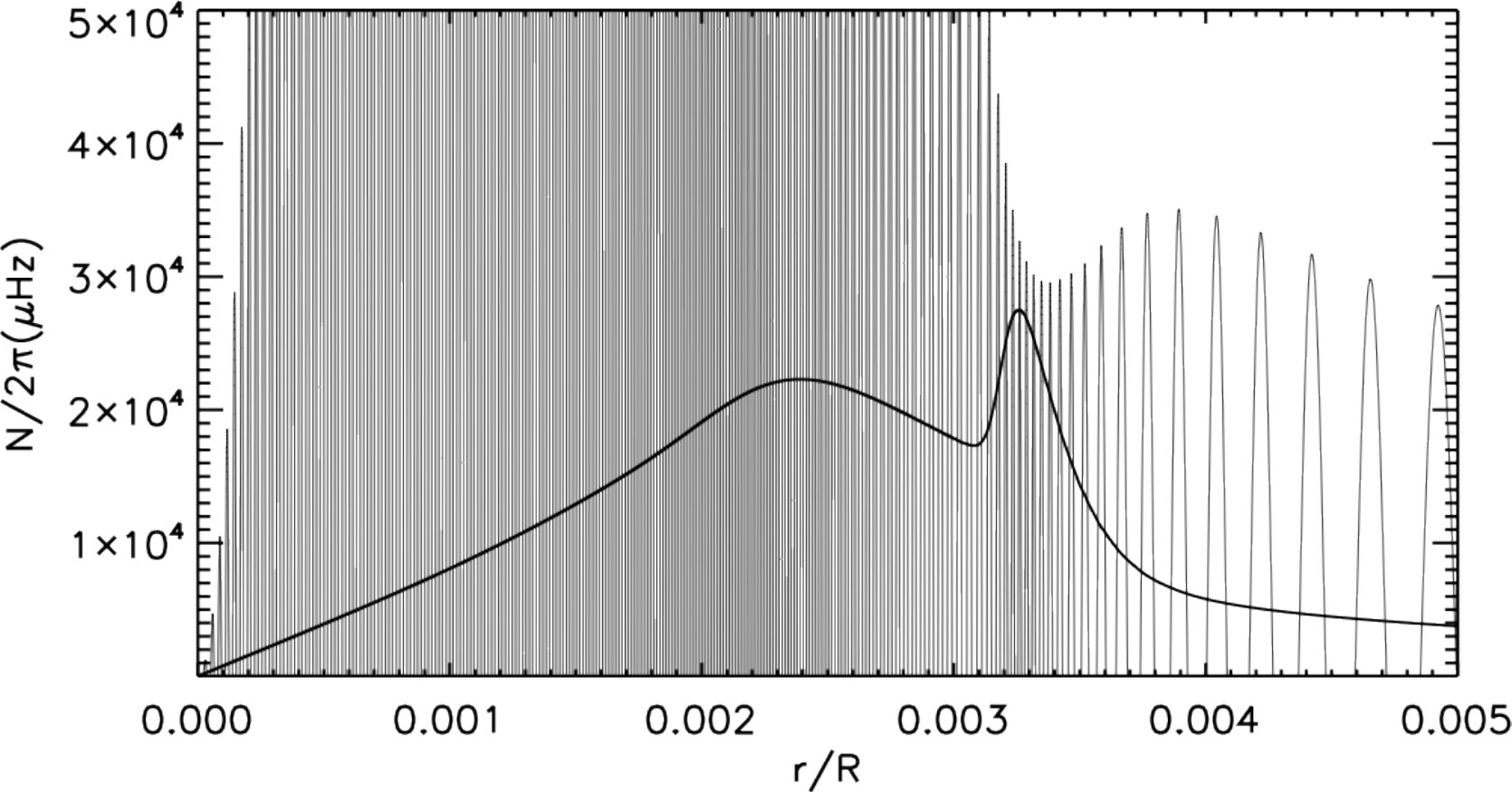}
\caption{Glitch (left) and no glitch (right).  Detail of the buoyancy frequency in a $M$=6M$_\odot$ main-sequence model (left, thick line) and a $M$=1M$_\odot$ RGB model (right, thick line). For comparison the eigenfunction $r^3\delta p$ multiplied by an arbitrary factor is show in each case (thin lines; where $\delta p$ is the Lagrangian pressure perturbation) for frequencies that may be observed in stars represented by these models, respectively,  $\nu=5.4\mu$Hz  and $\nu=51\mu$Hz  (see \cite{cunha19}, for further model details).}
\label{glitch}       % Give a unique label
\end{figure}

In Helio- and Asteroseismology, the term glitch usually refers to {\it a structural variation that is perceived as abrupt by the waves travelling across it}. Under this interpretation, a significant local variation in the structure of a star is a glitch only if it occurs within the wave's propagation cavity and the scale in which that variation takes place is comparable or smaller than the local wavelength of the mode under consideration.  Fig.~\ref{glitch} illustrates this concept based on two stellar models. On the left panel, the buoyancy frequency, $N$, in a 6 M$_\odot$ main-sequence (MS) model is seen to vary on a short scale at a fractional radius of $r/R=0.17$. This variation, which results from the chemical gradient built at the edge of the retracting convective core (see \cite{cunha19} for details) is perceived as a glitch by oscillations with frequencies that may be observed in such a star. On the other hand, on the right panel it is shown how $N$ varies in the inner layers of a red giant branch (RGB) model. A significant local variation of $N$ is also seen, at $r/R\approx 0.0033$, associated to the hydrogen-shell burning region, but because the modes that may be observed in this star have a rather short wavelength at that location, that variation does not constitute a glitch. 

\begin{figure}[t]
    \includegraphics[scale=.3]{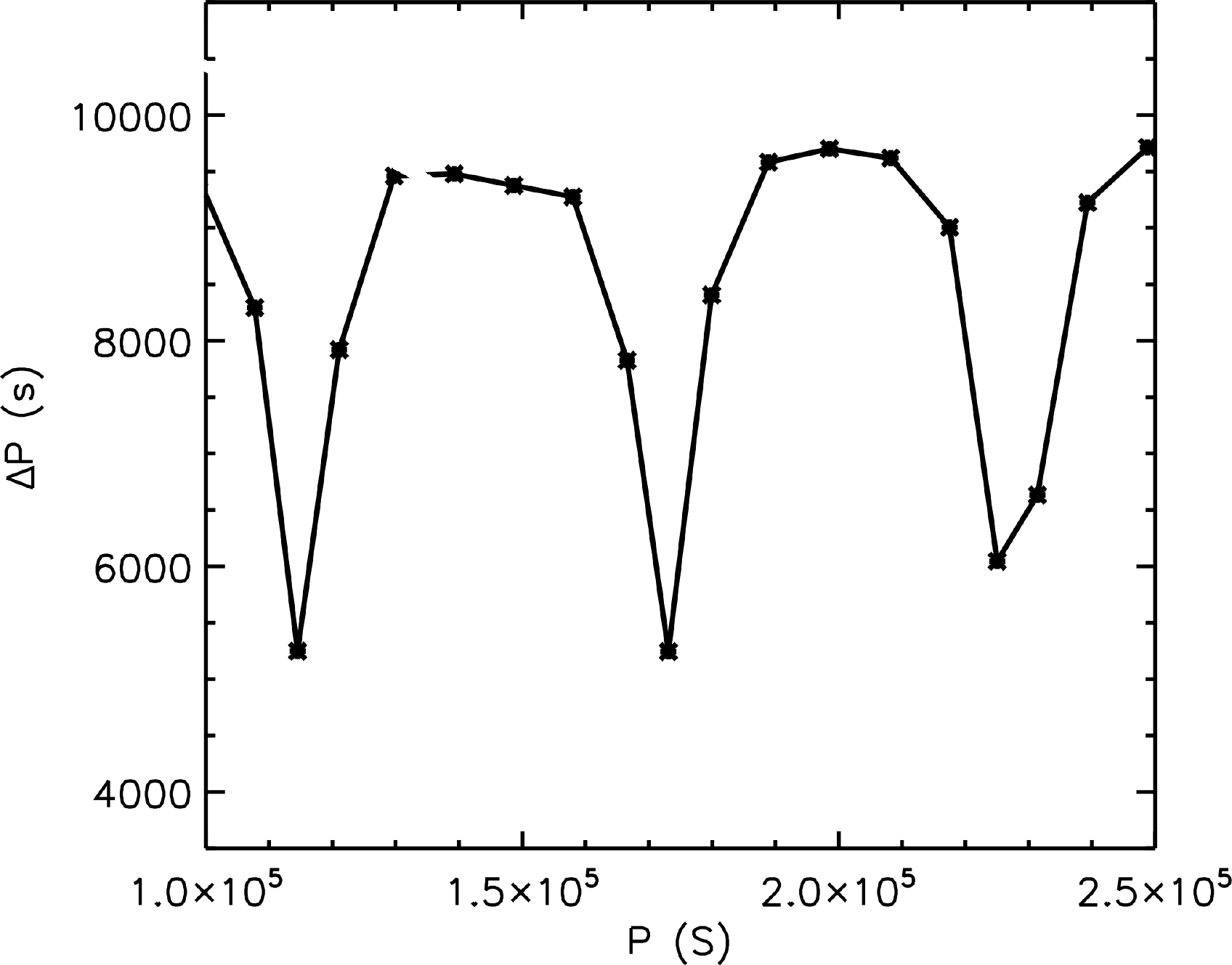}
	\includegraphics[scale=.38]{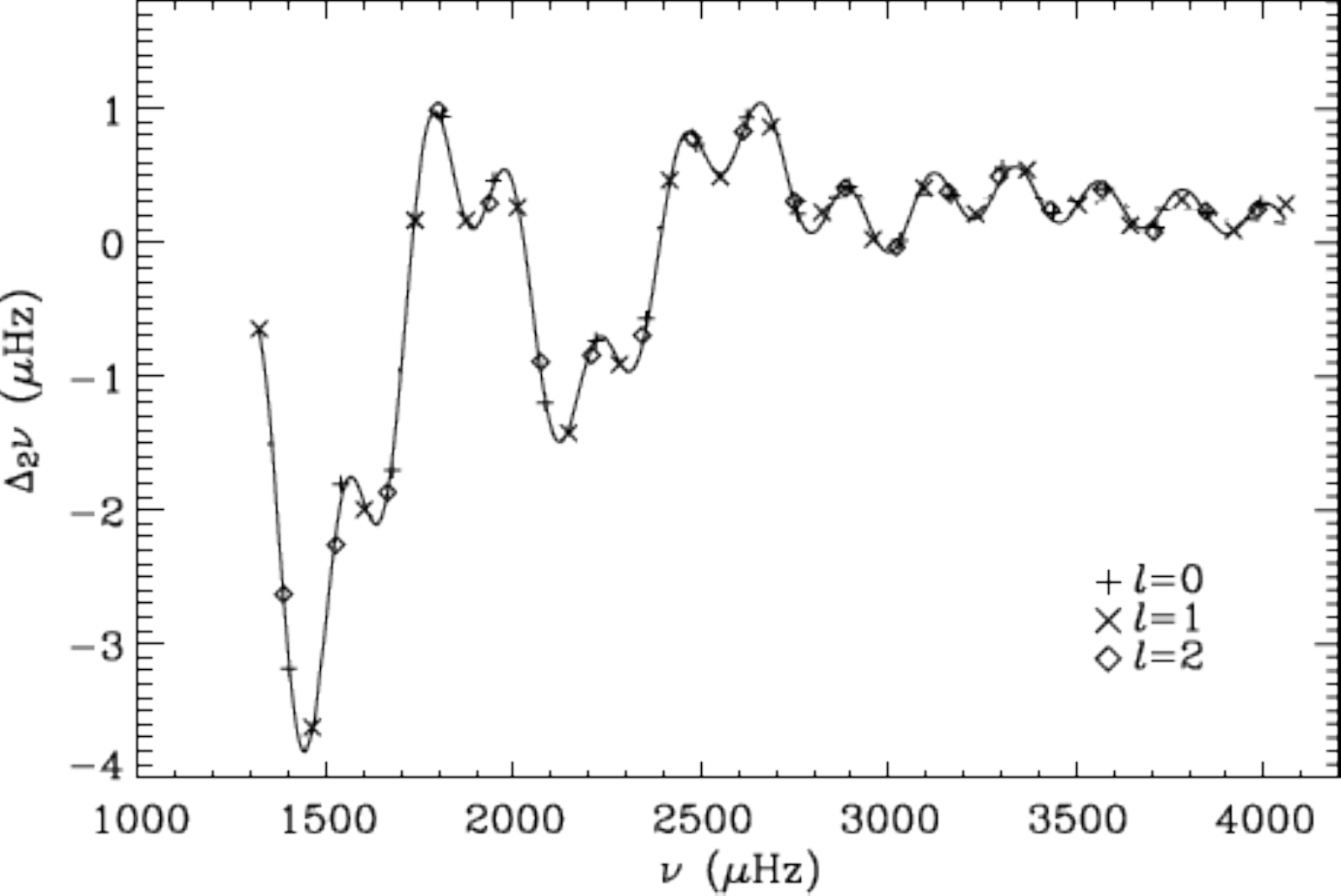}

	\caption{Left: Dipolar period spacings, $\Delta$P, for the MS model  shown in Fig.~\ref{glitch}. In the absence of a buoyancy glitch $\Delta$P would be approximately constant \cite{tassoul80}. Right: second frequency differences ($\Delta_2\nu$) \cite{gough90} for a model of the sun . The variations in $\Delta_2\nu$ result from the combined effect of the glitches at the base ofthe convective zone (variations on the shortest scale) and at the Helium I and II ionisation zones (variations on larger scales). Adapted from Houdek and Gough \cite{houdek07}, by permission of Oxford University Press on behalf of the Royal Astronomical Society.} 
	\label{signature}       % Give a unique label
\end{figure}

The phase of a wave is perturbed at a glitch and, as a result, the mode frequency is also perturbed with respect to the frequency it would have in the absence of the glitch. That perturbation shows a cyclic behaviour in frequency for pressure (p-) modes and in period for gravity (g-) modes. Perturbations to the frequencies resulting from the presence of a glitch are usually named {\it glitch signatures}. Examples of expected glitch signatures on the frequencies of g- and p-modes can be identified by computing the first period differences (or period spacings) and second frequency differences \cite{gough90}, respectively, as illustrated in Fig.~\ref{signature}.  The scale of the cyclic behaviour of the glitch signature provides information on the location of the glitch within the star, with glitches located closer to the center of the mode propagation cavity inducing signatures that vary on shorter scales than glitches located closer to the edges of the cavity  (e.g.,  \cite{houdek07, cunha15}). Moreover, the way the signature's amplitude varies with frequency depends on the structural quantity exhibiting the glitch, as well as on the shape and width of the glitch itself (e.g., \cite{monteiroetal94,cunha19}).
Finally, glitches can significantly change the energy distribution of the modes throughout the star, resulting in a concentration of mode energy in a particular region, a phenomena that is usually named  {\it mode trapping}.

Signatures of glitches are seen in stellar pulsators across the Hertzsprung–Russell (HR) diagram. Their diagnostic potential is unique, allowing us to probe directly localised features within the stars, whose origin and shape often depend on physical processes that are not fully understood or on the unknown abundance of a particular chemical element. Some early works attempting to make use of this diagnostic potential in the sun date back to the late 1980s, when Michael Thompson was still a PhD student \cite{GoughandThompson1988,thompson1988,vorontsov1988}. In these works, the authors discussed the signature that would be left on the solar frequencies by a  toroidal magnetic field confined at the base of the solar convective zone. While no signature of such a confined magnetic field was found on the data at that time, or in later data \cite{basu97}, the idea of looking for glitch signatures in stellar pulsations remained, with signatures of glitches from different physical origins being found and studied both in the sun and other stars over the following decades. In what follows I will summarise a selection of results from these studies.

\section{Acoustic glitches }
Glitches in the sound speed or sound-speed derivatives are usually termed {\it acoustic glitches}. The most common examples in solar-like stars are the glitch located at the base of the convective envelope, resulting from the rapid transition in the temperature gradient, and the glitch located at the helium second ionisation region, resulting from the drop in the first adiabatic exponent. Together they offer the opportunity to establish the depth of stellar convective envelopes, to improve our understanding of overshooting into the radiative regions, and to infer the abundance of helium, a key element for stellar evolution. This important diagnostic potential has been recognised by Michael Thompson and collaborators, as well as by other authors, in the 1990s, leading to a series of studies related to the inference of information from the acoustic glitch signatures seen in the solar data and to theoretical discussions of the potential impact of observing these signatures in other solar-like stars (e.g. \cite{gough90,houdek07,monteiroetal94,basu97,basu94,roxburgh94,monteiroetal00,basu04, monteiro05}). 

\begin{figure}[t]
	\includegraphics[scale=.415]{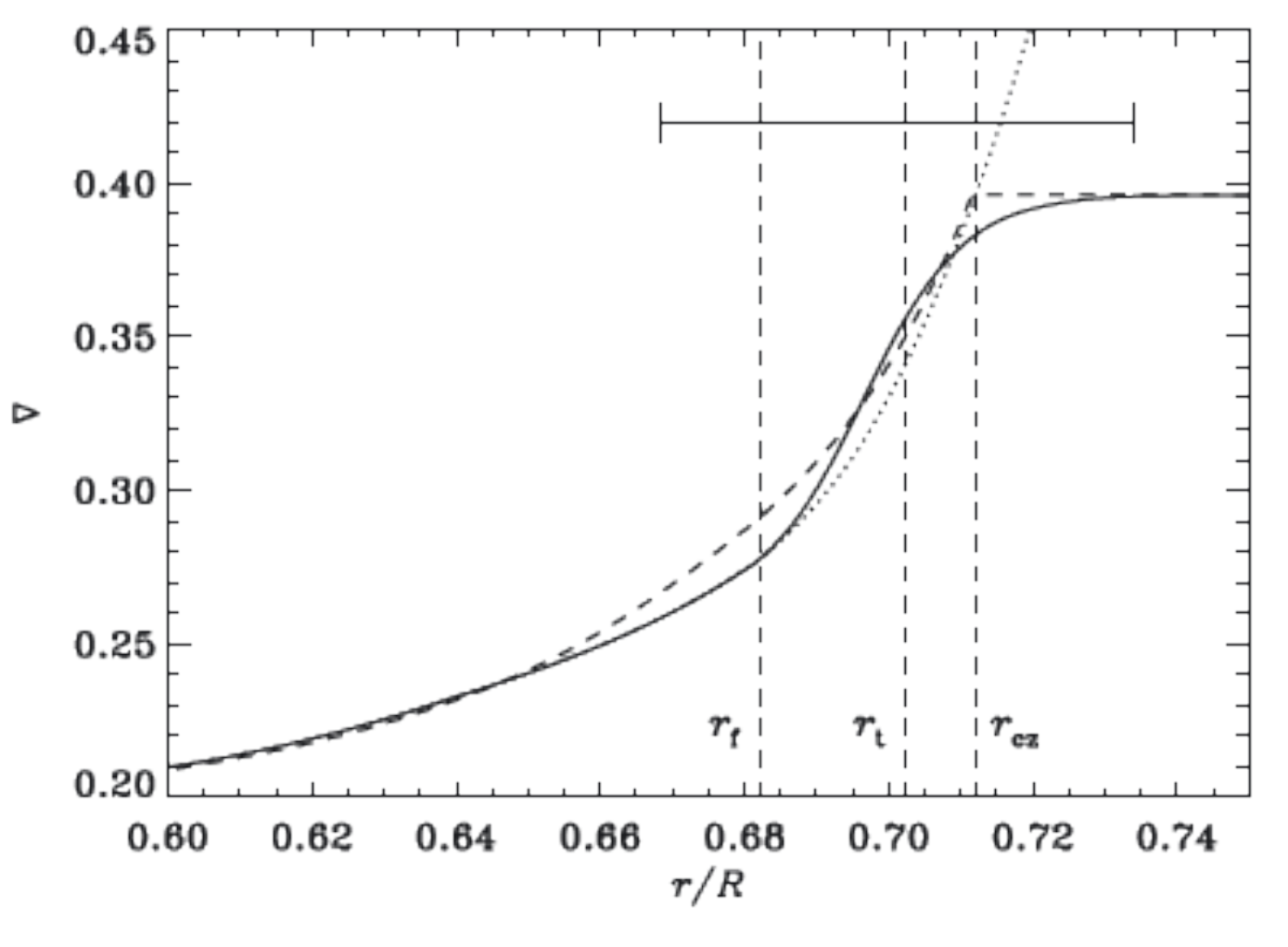}
	\includegraphics[scale=.38]{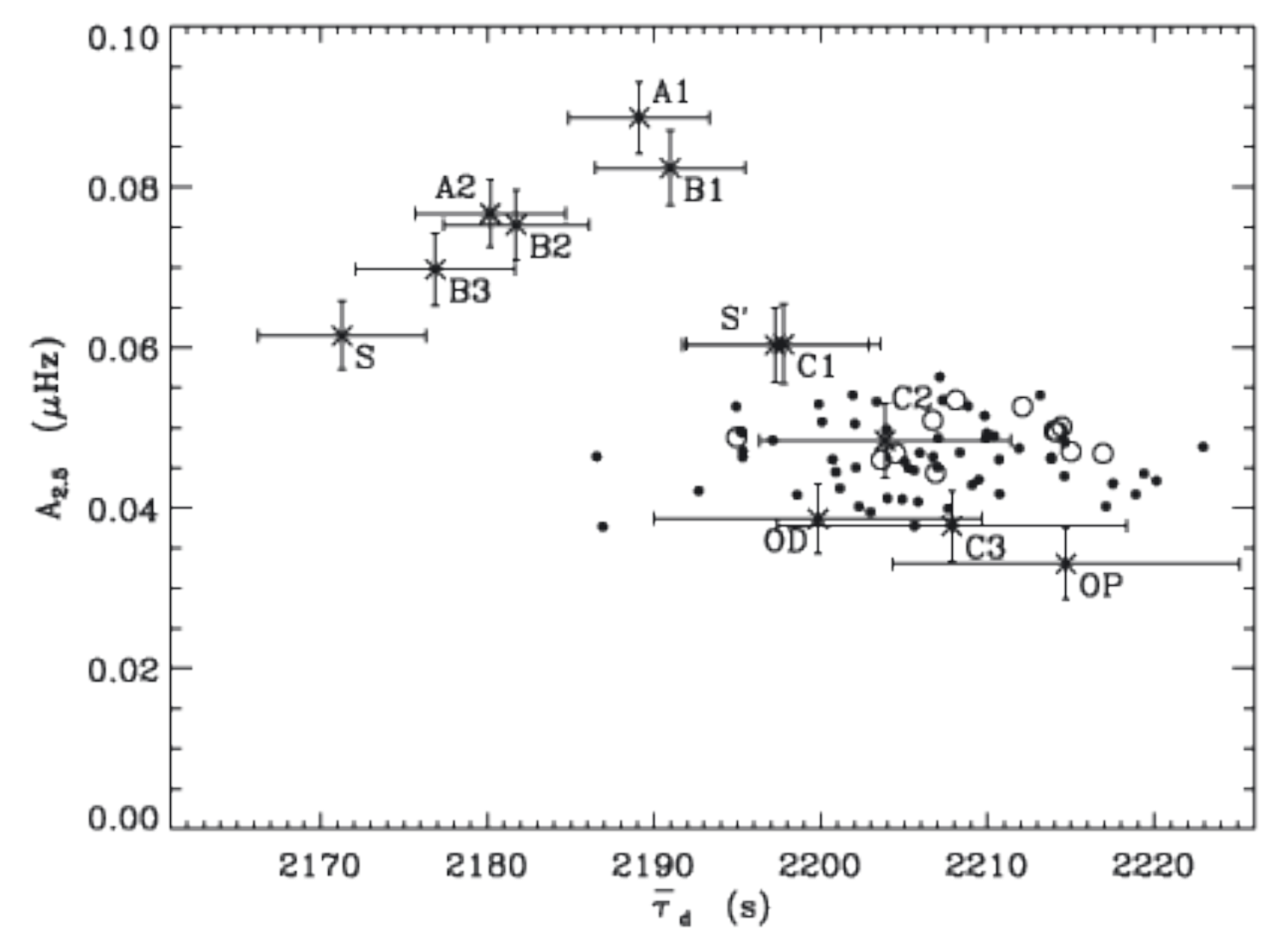}
		\caption{Left: Temperature gradients (d ln $T$/d ln $P$) for model S \cite{JCD96} without overshooting (dashed  curve) and for a second model (C2- continuous line) considered in \cite{JCD11a}, whose glitch signature is found to be consistent with that observed in the solar data. The dotted line is the radiative gradient
of model C2. Right: Comparison of two properties, $A_{2.5}$  and $\bar{\tau}$,  inferred from the glitch signatures of the series of models considered by \cite{JCD11a} (stars) and from the solar data (open and filled circles, derived from 1-yr and  72-d data sets, respectively). See  \cite{JCD11a} for details. Reproduced from \cite{JCD11a} by permission of Oxford University Press/on behalf of the Royal
Astronomical Society.}
	\label{jcd11}       % Give a unique label
\end{figure}

The signature from the base of a convective envelop depends critically on how overshoot modifies the temperature gradient at the transition. Earlier studies compared the glitch signature observed in the solar data with that derived from models with no overshoot, where the first derivative of the temperature gradient (hence, the second derivative of the sound speed) is discontinuous and that from models incorporating a simple overshoot prescription where the adiabatically stratified region is extended into the radiative region, leading to a discontinuity in the temperature gradient (hence, in the first derivative of the sound speed) \cite{monteiroetal94,basu94,roxburgh94}.  In all cases the glitch signature in the solar data was found to be comparable or smaller than that from models with no overshooting (hence, smaller that that from models including an overshoot region), favouring the models without overshooting and setting a maximum limit to the overshoot region of less than one tenth of the pressure scale height. However, the authors acknowledged that the adopted overshoot prescription was likely to be unrealistic and that a more meaningful picture than no overshooting would be one in which overshooting takes place, but where the transition in the temperature gradient at the base of the convective envelope is less abrupt. This possibility gained  support from a variety of overshooting models, including those derived from 3D numerical simulations, non-local convection and semi-analytic plum models, which seemed to allow for a significantly smoother transition (e.g., \cite{Brummell02,Rempel04,dengandxiong08}). With this possibility in mind \cite{JCD11a} compared the glitch signature from models with a range of possible temperature gradient profiles at the transition between the convective and radiative regions. An example of one such profile is shown by the continuous line on the left panel of Fig. \ref{jcd11}. On the right panel, the signature found in the solar data is compared with signatures derived  from the different model sequences considered in the same work. The comparison is based on $A_{2.5}$  which is related with the amplitude of the glitch signature at fixed frequency, and $\bar{\tau}$, related with the location of the glitch (see \cite{JCD11a} for precise definitions). The results show that the series of models with a smoother transition (series C) are more consistent with the solar data. The standard model S \cite{JCD96}, with no overshooting, as well as other models with sharp transitions (series A and B), all produce signatures with amplitudes that are significantly larger than the one found in the solar data. 

\begin{figure}[t]
%	\sidecaption[t]
\centering
    \includegraphics[scale=.4]{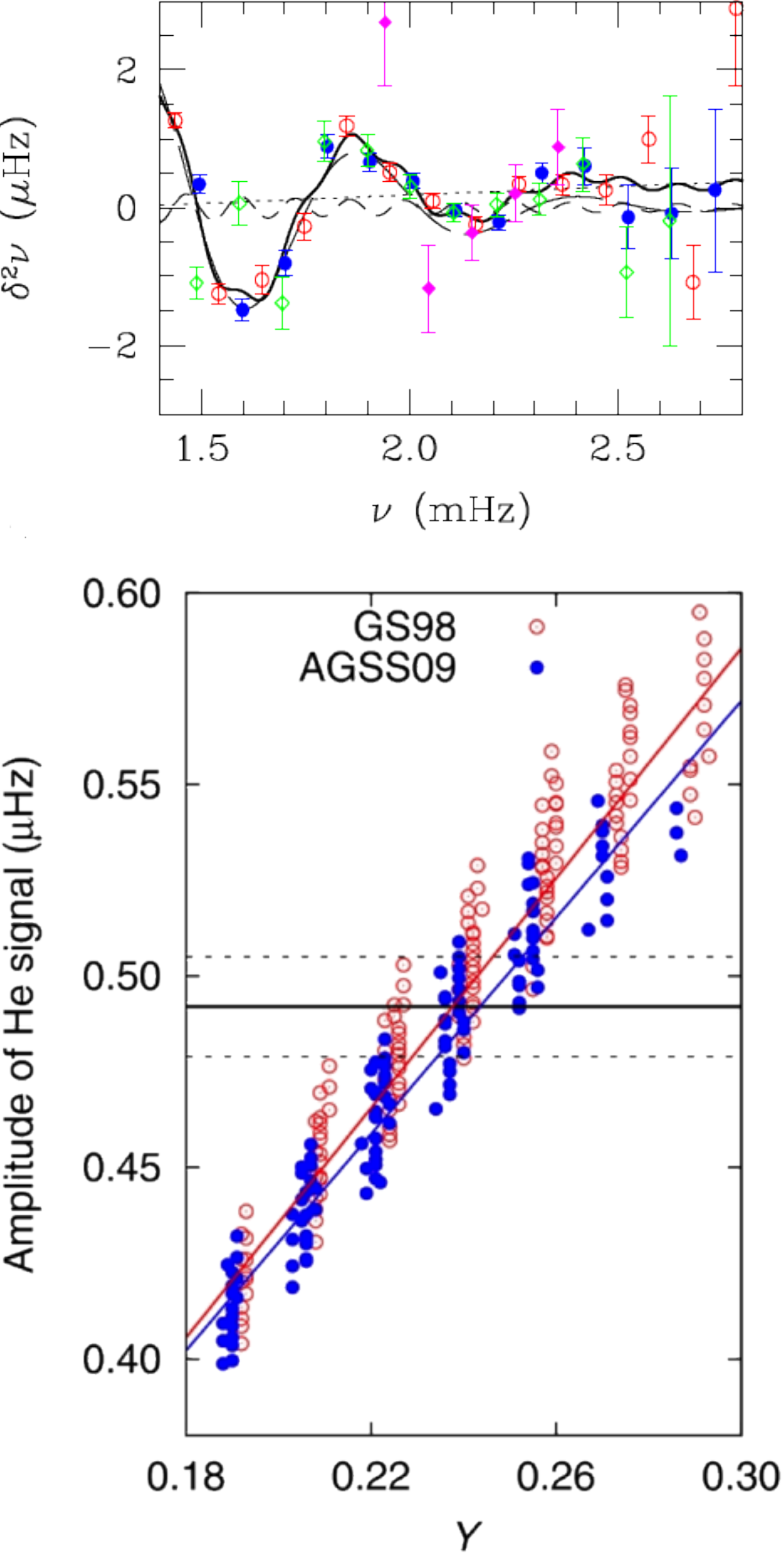}
    	\caption{Upper panel: Second frequency differences derived from data collected by the NASA {\it Kepler} satellite for the solar-like star 16 Cyg A (blue, red, green and magenta symbols for mode degrees  $l$=0,1,2 and 3, respectively).  The full line shows the fit to the data using the expected functional form of the signature seen in the data, while the long-dashed and dashed 
    	lines show the contributions to the  signature from the helium and the base of the convective envelope glitches, respectively. The dotted line is a smooth, non-glitch like component included in the fitted expression. 
    	 Bottom panel: comparison of the amplitude of the glitch signature inferred from the data (thick horizontal like with 1-$\sigma$ uncertainties given by the dashed lines) with the amplitudes derived from two model series with chemical mixtures from \cite{GS98}  (red) and \cite{AGSS09} (blue).  Figures from Verma et al. \cite{verma14}, \copyright AAS. Reproduced with permission. }
	\label{he1}       % Give a unique label
\end{figure}

\begin{figure}[t]
	\includegraphics[scale=.44]{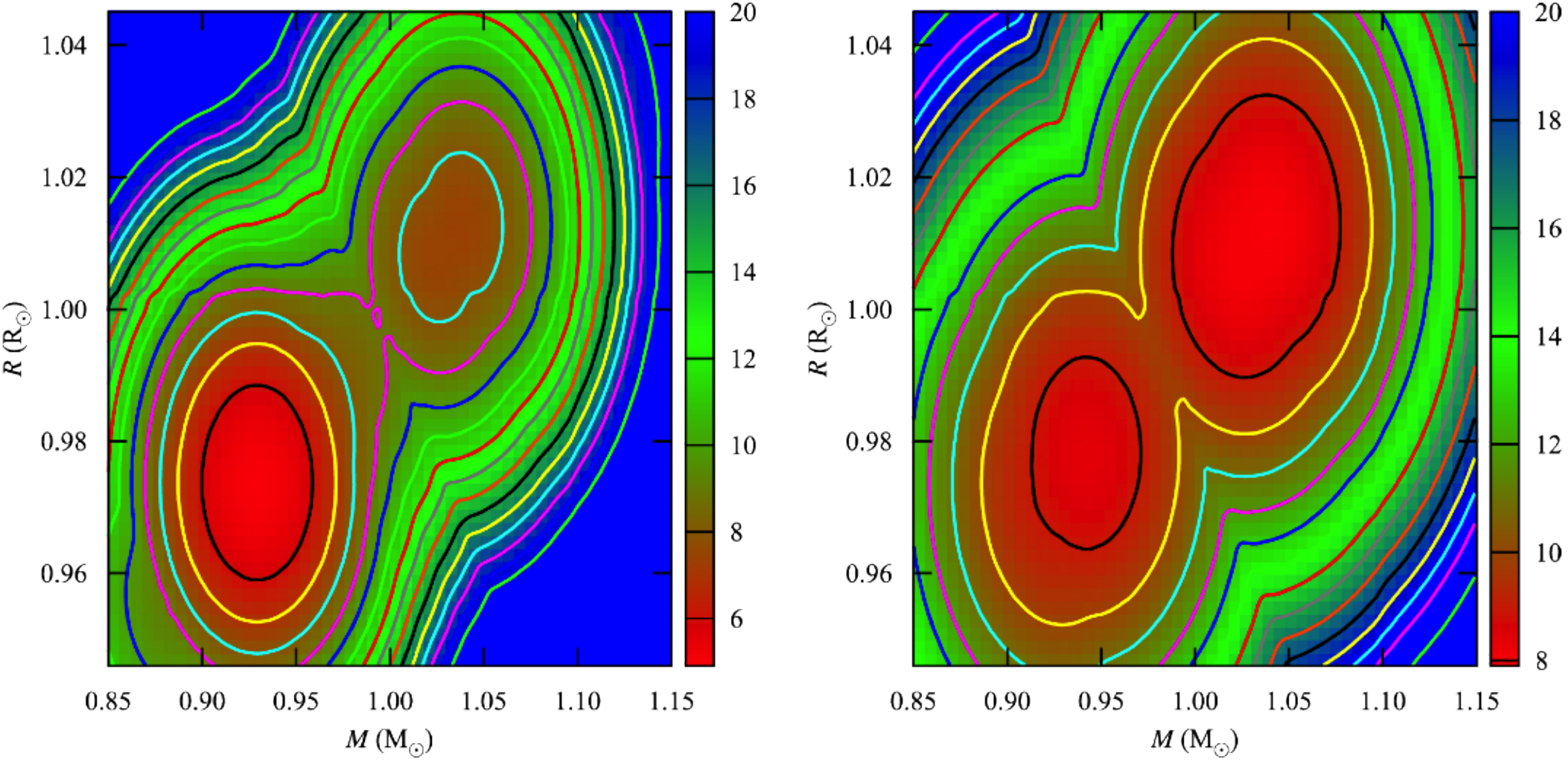}
	\caption{Chi-square maps for the parameters mass and radius derived from the forward modelling of data for the sun-as-a-star with noise levels similar to those found in stars from the {\it Kepler} legacy sample, where a lower chi-square (red) indicates a better fit to the data. The difference in the results shown in the two panels stems from the fact that the amplitude of the glitch signature was incorporated in the fitting procedure used on the right panel, but not in that used on the left panel.   Figure from Verma et al. \cite{verma17} \copyright AAS. Reproduced with permission.}
	\label{he2}       % Give a unique label
\end{figure}

With the advent of space-based data, the detection of glitch signatures from the base of convective envelopes in other solar-like stars became possible \cite{mazumdar12,mazumdar14,verma17}. From studying these signatures it is possible to establish how the extent of stellar convective envelopes depends on stellar properties such as mass, age, and metallicity, further constraining stellar evolution and stellar dynamo theory.  Space-based data offers also the opportunity to constrain the helium abundance is solar-like stars, where it cannot be directly determined from the analysis of spectroscopic data. This element has a significant influence on how a star evolves. The fact that its  abundance is only weakly constrained (usually based of galactic evolution models) thus leads to significant uncertainties in the evolution path of a star and, due to the mass-helium degeneracy in main-sequence stellar evolution, to large uncertainties in the determination of the stellar mass, given a large uncertainty in the helium abundance. The opportunity to constrain the helium abundance directly through the study of the glitch signature associated to the depression in the adiabatic exponent  at the location of the helium second ionisation is, therefore, a very interesting one. The underlying idea is that the amplitude of the helium glitch signature is predominantly a function of the helium abundance in the helium ionisation zone and, thus, can be calibrated against models where the helium abundance in known.  An example of this was provided by \cite{verma14}, where the authors used the amplitude of the helium glitch signatures derived from the oscillations of the binary system 16Cyg A and B to infer the current helium abundance in their envelopes. The results are illustrated in Fig.~\ref{he1} for the case of  16Cyg A, reproduced from their work, the upper panel showing the fit to the glitch signature derived from the observations and the lower panel showing the star-models comparison of the amplitudes. The spred in the model results at fixed helium is mostly associated with differences in other model properties, such as mass, effective temperature, and luminosity (same colour) or  in the chemical mixture considered (different colours). Despite this spread, it is clear that the amplitude of the glitch signature is very strongly dependent of the helium abundance, allowing the authors to derive the envelope helium abundance in both stars. The work was later extended to the study of the {\it Kepler} legacy sample \cite{verma19}. Interestingly, in \cite{verma17} the authors have shown that incorporating the information inferred on the glitch amplitude in the forward modelling of stars can help removing biases in the inferred stellar parameters resulting from the helium-mass degeneracy. This is illustrated in Fig.~\ref{he2} reproduced from \cite{verma17} which shows the Chi-squared maps of the stellar parameters mass and radius resulting from two different forward modelling attempts of the data for the sun-as-a-star prepared by \cite{lund17} (where the noise level was made similar to that of the data for the {\it Kepler} legacy sample stars).  The left panel shows the results from the modelling conducted without considering the information on the amplitude of the signature of the helium glitch, while the right panel shows the results when that information is incorporated explicitly. In the absence of the glitch information a lower radius and mass are preferred. However, the glitch amplitude  adds information on the helium abundance content, leading to results that are in better agreement with the solar values. 
 
\section{Buoyancy glitches}
Buoyancy glitches refer to glitches in the buoyancy  frequency, $N$. They affect the pulsation properties of g modes and mixed modes and are, thus, important in the context of g-mode pulsators, such as main-sequence intermediate-mass, Subdwarf B, and white dwarf stars, and mixed-mode pulsators, such as red giants, both during the RGB ascent and during the helium-core-burning (HeCB) phase.  Through the analysis of buoyancy glitches it is possible to address key questions in the physics of g-mode and mixed-mode pulsators, related to the chemical stratification in compact pulsators \cite{brassard92} and to the mixing conditions at the edge of convective cores  in core-convective hydrogen-burning stars and in HeCB stars \cite{bossini15}. 

\begin{figure}[t]
%	\sidecaption[t]
\centering
    \includegraphics[scale=1]{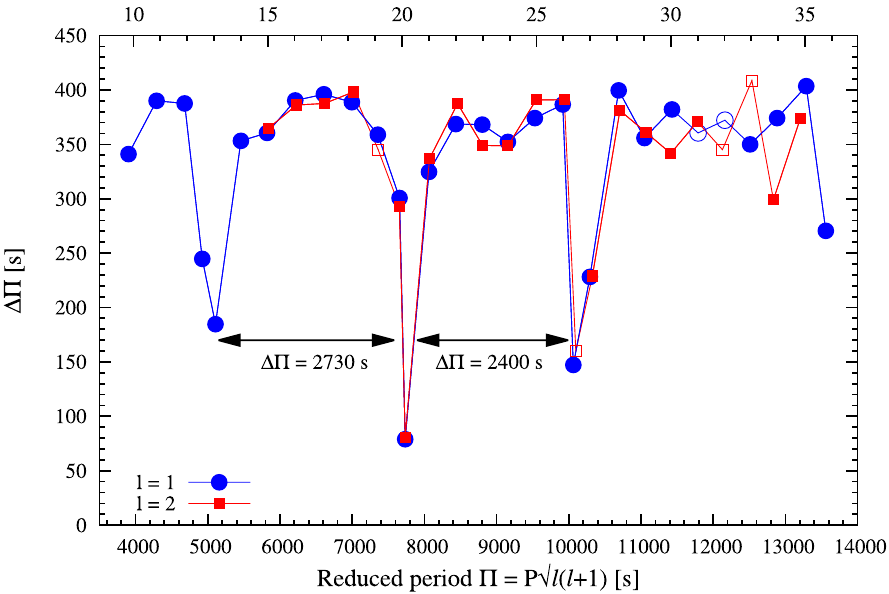}
	\caption{Reduced period spacings computed from modes of degree $l=1$ (blue) and degree $l=2$ (red) for the subdwarf B star KIC10553698A  where the reduced period is defined by $\Pi=P\sqrt{l(l+1)}$. The deviation from a constant value is induced by a glitch in $N$. Figure from  \cite{ostensen14}, reproduced with permission, \copyright ESO.}
	\label{subd}       % Give a unique label
\end{figure}

\begin{figure}[t]
%	\includegraphics[scale=.33]{Ostensen2014_fig10.pdf}
%	\sidecaption[t]
\centering
	\includegraphics[scale=.5]{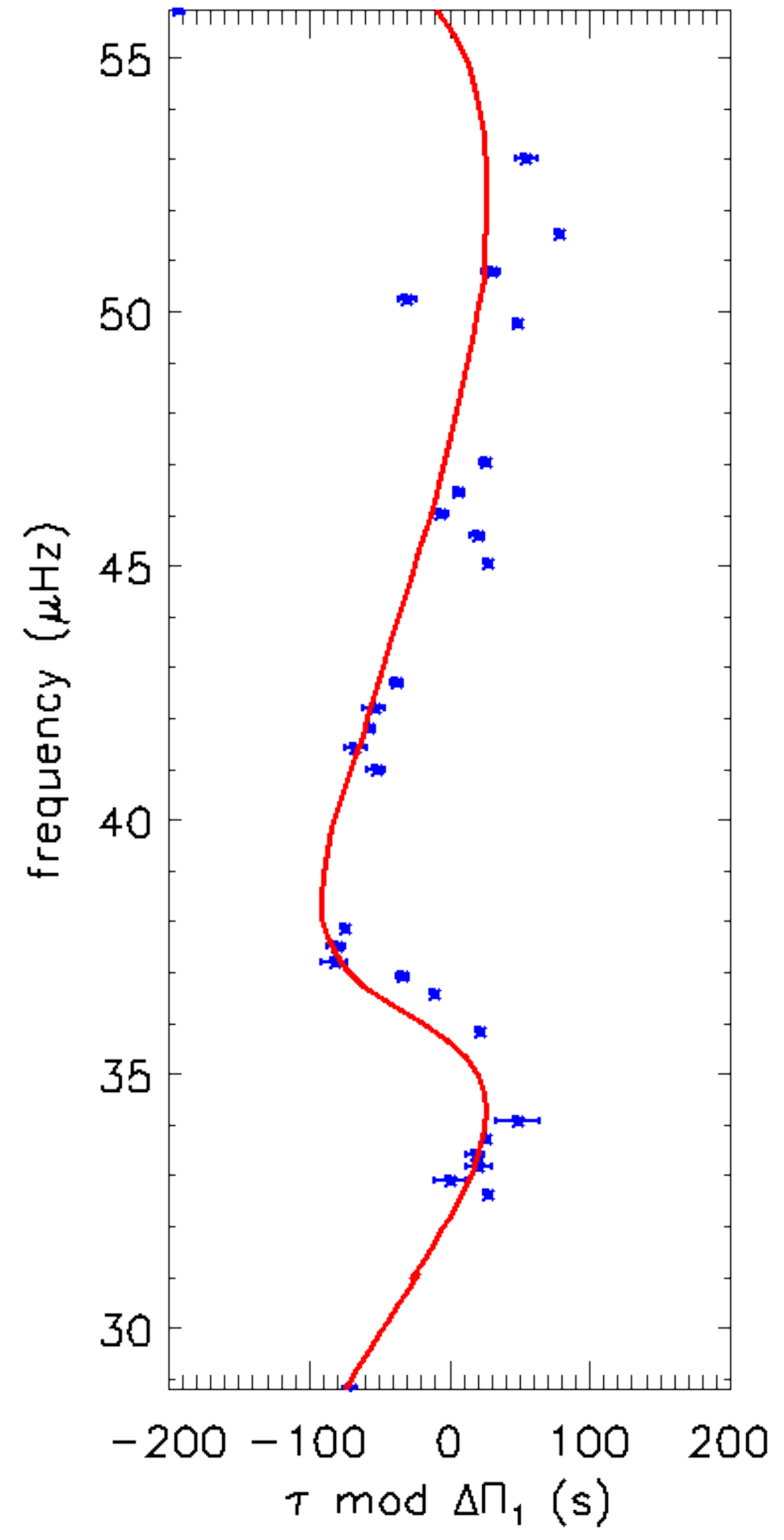}
	\caption{Stretched period echelle diagram for the HeCB red-giant star KIC 009332840. The stretched  periods correspond to the g-mode  periods that would be observed if no coupling existed between the gravity and acoustic waves (see \cite{mosser15} for an exact definition). They would be constantly spaced if no glitch was present (in a spherically symmetric star).  The deviation of the blue symbols (representing the data) from a vertical alignment is induced by a glitch in $N$. The red line shows the best fit of the asymptotic expression derived taking into account the glitch effect \cite{cunha19} to the data.  Figure from \cite{vrard19}.}
	\label{clump}       % Give a unique label
\end{figure}

Amongst the most promising recent results in this  context is the detection of glitch signatures in HeCB stars, both in red giants \cite{mosser15,vrard19} and in subdwarf B stars \cite{ostensen14,ghasemi17}, which are thought to hold information on the chemical stratification in the innermost layers of these stars. That stratification depends on the details of the mixing beyond the helium-burning core, which is largely unknown and represents one of the key open questions in present-day stellar astrophysics \cite{bossini15}.  Figs~\ref{subd}-\ref{clump} show examples of glitch signatures found in two HeCB stars, namely, a subdwarf B star and a red-giant star, respectively. Ideally, one would hope to infer the information contained in stellar glitch signatures in a model independent way. This is even more important in the case of the buoyancy glitches discussed here, as the models of stars in the HeCB phase are particularly complex and, at the same time, less robust and less tested than those of low-mass main-sequence stars.  An important step towards this goal is the incorporation of the glitch effect in the asymptotic analysis of the frequencies (or periods) of g- and mixed modes. The glitch-induced perturbation to the periods of g modes has been theoretically addressed in the context of wite dwarfs and main-sequence intermediate-mass stars in the past \cite{brassard92, miglio08}. Recently, analytical expressions describing directly the behaviour of the period spacings in the presence of buoyancy glitches of different amplitudes and functional forms in pure g-mode pulsators were presented by \cite{cunha19}. The authors also extended the work by \cite{cunha15}, where the first asymptotically-based analytical expression for period spacings of mixed modes in the presence of buoyancy glitches had been presented, by considering different possible functional forms for the glitches in red-giant pulsators. An exemple of the fit of the analytical expression derived by  \cite{cunha19} to the data on a HeCB red-giant star is shown by the red curve in Fig.~\ref{clump}. Through the fit, the authors were able to infer the location and amplitude of the glitch in the red-giant star \cite{vrard19}.

\section{Concluding remarks}
The study of glitches allows us to probe specific layers within stars whose structural details are often shaped by physical processes that are most challenging to model. Moreover, it allows us to infer information on the abundance of helium, a key element for stellar evolution that cannot otherwise be constrained in cool stars.  As a consequence, the study of glitches offers a unique potential to further advance our  understanding of stellar structure, dynamics, and evolution. Its foundation was established in the early days of helioseismology. Michael Thompson has undoubtedly understood the potential of this technique and played a major role in its development, contributing, along with a few others in the field, to set the first stones that paved the way for it to be more generally applied. Following on those earlier developments, and also on developments of similar techniques in the context of the study of g modes in white dwarfs, the study of glitches has recently gained a new breath, thrusted by the exquisit space-born long asteroseismic data sets made available the NASA satellite {\it Kepler}, and shall continue to develop into the future, in particular with the expected launch of the PLATO mission, by ESA.

\begin{acknowledgement}
The author would like to express her profound gratitude to the Scientific Organizing Committee for the opportunity to speak at the conference honouring the life and work of Michael Thompson. 
	MSC is supported in the form of a work contract funded through Fundação para a Ciência e Tecnologia (FCT). This work was supported by FCT through the research grants UID/FIS/04434/2019, UIDB/04434/2020 and UIDP/04434/2020 and by FCT through national funds (PTDC/FIS-AST/30389/2017) and by FEDER - Fundo Europeu de Desenvolvimento Regional through COMPETE2020 - Programa Operacional Competitividade e Internacionalização (POCI-01-0145-FEDER-030389).
\end{acknowledgement}

%%%%%%%%%%%%%%%%%%%% REFERENCES %%%%%%%%%%%%%%%%%%

% The best way to enter references is to use BibTeX:

\bibliographystyle{unsrt}
%\bibliography{solar-like_v1} % if your bibtex file is called example.bib
\bibliography{mscunha} % if your bibtex file is called solar-like.bib

%%%%%%%%%%%%%%%%%%%%%%%%%%%%%%%%%%%%%%%%%%%%%%%%%%

\end{document}